 \documentclass[preprint2]{aastex}
\usepackage{graphics,epsfig,longtable,lscape,txfonts,color}
\usepackage[english]{babel}

\newcommand{\myemail}{hstiele@mx.nthu.edu.tw}
\def\deg{\hbox{$^\circ$}}
\newcommand{\mr}{\mathrm}
\newcommand{\nh}{\hbox{$N_{\rm H}$}}
\newcommand{\hcm}[1]{$\times 10^{#1}$ cm$^{-2}$}

\newcommand{\oergcm}[1]{$10^{#1}$~erg~cm$^{-2}$~s$^{-1}$}

\def\ie{i.\,e.}                                      
\def\eg{e.\,g.}                                      

\def\xmm{\textit{XMM-Newton}}
\def\swift{\textit{Swift}}
\def\nus{\textit{NuSTAR}}

\def\gx339{GX\,339--4}
\def\h1743{H\,1743--322}

\def\sw19{Swift\,J1910.2--0546}

\def\ga{\mathrel{\mathchoice {\vcenter{\offinterlineskip\halign{\hfil
$\displaystyle##$\hfil\cr>\cr\sim\cr}}}
{\vcenter{\offinterlineskip\halign{\hfil$\textstyle##$\hfil\cr
>\cr\sim\cr}}}
{\vcenter{\offinterlineskip\halign{\hfil$\scriptstyle##$\hfil\cr
>\cr\sim\cr}}}
{\vcenter{\offinterlineskip\halign{\hfil$\scriptscriptstyle##$\hfil\cr
>\cr\sim\cr}}}}}                                            

\shorttitle{2015 outburst decay of \gx339}
\shortauthors{Stiele, Kong}

\begin{document}


\title{\nus\ and \xmm\ observations of the 2015 outburst decay of \gx339}

\author{H. Stiele}
\affil{National Tsing Hua University, Department of Physics and Institute of Astronomy, No.~101 Sect.~2 Kuang-Fu Road,  30013, Hsinchu, Taiwan}
\email{\myemail}
\and

\author{A. K. H. Kong}
\affil{National Tsing Hua University, Department of Physics and Institute of Astronomy, No.~101 Sect.~2 Kuang-Fu Road,  30013, Hsinchu, Taiwan}
\affil{Astrophysics, Department of Physics, University of Oxford, Keble Road, Oxford OX1 3RH, UK}





\begin{abstract}
The extent of the accretion disk in the low/hard state of stellar mass black hole X-ray binaries remains an open question. There are some evidence suggesting that the inner accretion disk is truncated and replaced by a hot flow, while the detection of relativistic broadened iron emission lines seems to require an accretion disk extending fully to the innermost stable circular orbit.
We present comprehensive spectral and timing analyses of six \textit{Nuclear Spectroscopic Telescope Array} (\nus) and \xmm\ observations of \gx339\ taken during outburst decay during the autumn of 2015. Using a spectral model consisting of a thermal accretion disk, Comptonized emission, and a relativistic reflection component we obtain a decreasing photon index, consistent with an X-ray binary during outburst decay.
Although, we observe a discrepancy in the inner radius of the accretion disk and that of the reflector, which can be addressed to the different underlying assumptions in each model, both model components indicate a truncated accretion disk that resiles with decreasing luminosity. The evolution of the characteristic frequency in Fourier power spectra and their missing energy dependence support the interpretation of a truncated and evolving disk in the hard state.
The \xmm\ dataset allowed us to study, for the first time, the evolution of the covariance spectra and ratio during outburst decay.  The covariance ratio increases and steeps during outburst decay, consistent with increased disk instabilities.  
\end{abstract}

\keywords{X-rays: binaries -- X-rays: individual: \gx339\ -- binaries: close -- stars: black hole}

\section{Introduction}
\gx339\ can be regarded as the archetypical low-mass black hole X-ray binary. Since its discovery in 1973 by the OSO-7 satellite \citep{1973ApJ...184L..67M}, it showed frequent outbursts of varying strength, and there is a possible $\sim$200 days long-term variability \citep{2002MNRAS.329..588K}. Both, the mass of the black hole in and the distance to \gx339, are rather uncertain, estimated at $M\simeq10 M_{\odot}$ and $d\simeq8$ kpc by \citet{2004MNRAS.351..791Z}. Based on \nus\ observations, \citet{2016ApJ...821L...6P} used the spin and inclination they determined from modelling the reflection component as input for the continuum fitting to obtain a mass of $9.0^{+1.6}_{-1.2} M_{\odot}$ and a distance of $8.4\pm0.9$ kpc just from the X-ray spectrum.

During most of its outbursts \gx339\ evolves from the so called low/hard state (LHS) through the hard and soft intermediate states (HIMS/SIMS) to the high/soft state (HSS), where it can remain for several weeks, before it returns at lower luminosity into the LHS, passing again through the SIMS and HIMS. Here, we follow the classification of \citet{2010LNP...794...53B}; see however \citet{2006csxs.book..157M} for an alternative classification. The different states through which the black hole transient evolves can be identified in the hardness intensity diagram \citep[HID;][]{2005A&A...440..207B,2005Ap&SS.300..107H,2006MNRAS.370..837G,2006csxs.book..157M,2009MNRAS.396.1370F,2010LNP...794...53B}, as the hard and soft states have distinct spectral properties. In the LHS the X-ray spectrum is dominated by Comptonized emission, which can be described by a power law with a photon index of $\Gamma \le 1.7$ and cut-off energies of $\sim$ 50 -- 100 keV. In the HSS the X-ray spectrum is dominated by thermal emission of the accretion disk. Furthermore, hard and soft states show distinct variability properties. The power density spectra (PDS) of the LHS show strong (up to 40 -- 50 per cent) band limited noise (BLN) and quasi-periodic oscillations (QPOs), while PDS of the HSS show weak (few per cent) power law noise.     

While there is strong observational evidence that for a large fraction of the HSS the accretion disk extends all the way down to the innermost stable circular orbit \citep[ISCO;][]{2004MNRAS.347..885G,2010ApJ...718L.117S}, the accretion geometry in the LHS has been a subject of intense debate in recent years. In the quiescent state, the disk is found to recede \citep{1995ApJ...442..358M,2001ApJ...555..477M,2003ApJ...593..435M,1995ApJ...452..710N,1996ApJ...457..821N,1997ApJ...489..865E,2001ApJ...555..483E}. This suggests that a truncated disk is required in early stages of an outburst, and that the LHS is dominated by emission of an optically thin, advection-dominated accretion flow \citep[see \eg\ ][]{1997ApJ...489..865E} surrounded by an accretion disk truncated far away from the ISCO \citep{2007A&ARv..15....1D,2010LNP...794...53B,2014ARA&A..52..529Y}.  During the outburst, the inner accretion disk must then evolve and extend itself closer to the black hole.

Based on observations of soft emission in the LHS, which is often attributed to disk emission, the inner disk radius can be determined by modelling this component. Alternatively, the disk radius can be inferred from modelling the reflection component, as the disk acts as a reflector. \gx339\ is a key source in the inner disk truncation debate. Some studies of \gx339\ at luminosities $\ge1$\% $L_{\mr{Edd}}$ suggested an inner disk radius consistent with a disk extending down to the ISCO \citep{2006ApJ...653..525M,2008ApJ...680..593T,2014A&A...564A..37P}. However, \citet{2010MNRAS.407.2287D} using the same data as \citet{2006ApJ...653..525M} and \citet{2008ApJ...680..593T}, found a large truncation radius. The discrepancy was attributed  to the severe pile-up in the \xmm\ MOS data, which broadens the iron line. A highly truncated disk has been also found by \citet{2014MNRAS.437..316K}, who calculated the inner radius both from modelling the disk emission and that of the reflection component. Interestingly, the radii calculated by the two methods do not agree with each other. A study of the 2013 outburst, in which \gx339\ always remained in the LHS, found an inner disk radius of $R_{\mr{in}}\sim20R_{\mr{g}}$ from both disk and reflection spectral components \citep{2014arXiv1411.7411P}. Based on a systematic study of the iron line region, that tracked the evolution of the inner  accretion disk in the LHS, \citet{2015A&A...573A.120P} found an accretion disk that extended closer to the black hole at higher luminosities, but was consistent with being truncated throughout the entire LHS. \citet{2016MNRAS.458.2199B} reanalysed seven archival \xmm\ observations of \gx339\ in the hard state at outburst rise, fitting the data with a modified \texttt{diskbb} \citep{1984PASJ...36..741M} plus \texttt{relxill} \citep{2014MNRAS.444L.100D,2014ApJ...782...76G} model (they split the normalisation of the \texttt{diskbb} component into two factors, one depending on the mass and distance, and one depending on the `true' inner disk radius and inclination, where the latter one can then be linked to the corresponding parameters of the \texttt{relxill} model). They found that the inner disk radius of the accretion disk could not be set equal to that of the reflector and obtained an evolution of the inner disk radius consistent with the truncation disk model. The discrepancy between the two radii implied that the soft spectral component was not a standard blackbody disk. Studies of \gx339\ at luminosities below 1\% $L_{\mr{Edd}}$ found a truncated accretion disk \citep{2009ApJ...707L..87T,2013ApJ...774..135A}.

Reflection modelling and disk continuum models also allow us to estimate the spin and inclination. Although there is a discrepancy between the results obtained with these two methods, which is likely due to different underlaying assumptions in the models \citep{2014SSRv..183..277R,2014SSRv..183..295M}, a Schwarzschild (\ie\ non-spinning) black hole can be ruled out with high significance \citep{2008ApJ...679L.113M,2010MNRAS.406.2206K}. Most studies prefer a spin of $a>0.9$ \citep[see][for a recent overview and references therein]{2016ApJ...821L...6P}. Only \citet{2010MNRAS.406.2206K} argued, based on continuum modelling and assuming an inclination $>45$\deg, that the spin should be less than 0.9. 

The inclination of \gx339\ is only weakly constrained by dynamic measurements. It must be less than 60\deg, as the system is non-eclipsing \citep{2002AJ....123.1741C}, and a plausible lower limit of $i>45$\deg\ is given from the mass function \citep{2004MNRAS.351..791Z}. Measurements of the inner disk inclination from the broad iron line prefer rather low values: $i=12$\deg$^{+4}_{-2}$ \citep{2004ApJ...606L.131M}, $i=19^\circ\pm1^\circ$ \citep{2008ApJ...679L.113M}, $i=18.\!\!^\circ2^{+0.3}_{-0.5}$ \citep{2008MNRAS.387.1489R}, while using relativistic reflection models give values $\ge 30$\deg: $36^\circ\pm4^\circ$ \citep{2014ApJ...782...76G}, $48^\circ\pm1^\circ$ \citep{2015ApJ...813...84G}, 31\deg to 59\deg \citep[depending on the model;][]{2015ApJ...806..262L}, $30^\circ\pm1^\circ$ \citep{2016ApJ...821L...6P}. A combined spectroscopy and timing analysis fitting both the lag and spectral data obtained $i<30$\deg \citep{2012MNRAS.422.2407C}.

By fitting mean X-ray spectra, one studies the time-averaged spectral shape of a source, learning nothing about how the individual spectral components vary with respect to each other in time. One technique to construct `variability spectra' is to obtain a PDS for each individual energy channel and to integrate the PDS over a given frequency range to measure the variance in that channel. That way one can construct an rms spectrum, and examine the components which vary over that frequency range \citep[\eg\ ][]{1999A&A...347L..23R,2003MNRAS.345.1271V}.
\citet{2009MNRAS.397..666W} developed a technique called covariance spectra, which is similar to rms spectra and allows to disentangle the contribution of spectral components to variations on different time-scales. The covariance is derived between the channel of interest and a broader reference band, which makes it more robust regarding low signal-to-noise data.
Covariance spectra and ratios derived from \xmm\ data taken during the LHS at the beginning of an outburst revealed an increased disk blackbody variability with respect to the Comptonized emission below 1 keV at time scales longer than 1 s \citep{2009MNRAS.397..666W,2012MNRAS.427.2985C}.  On shorter timescales variability of the Comptonized component drives the disk variability, consistent with propagating models modified by disk heating at short timescales.

In this paper, we present a comprehensive study of the spectral and temporal variability properties of six \nus\ and \xmm\ observations of \gx339\ taken during decay of its 2014/15 outburst.

\begin{table*}
\caption{Details of \xmm\ observations}
\begin{center}
\begin{tabular}{lrlrrrr}
\hline\noalign{\smallskip}
 \multicolumn{1}{c}{\#} & \multicolumn{1}{c}{Obs.~id.} & \multicolumn{1}{c}{Date}  & \multicolumn{1}{c}{Mode$^{\ddagger}$} &  \multicolumn{1}{c}{Net Exp. [ks]}  &  \multicolumn{1}{c}{Exp.$^{\dagger}$ [ks]}  &  \multicolumn{1}{c}{region$^{*}$} \\
 \hline\noalign{\smallskip}
1 & 0760646201  & 2015-08-28 & T & 14.9 & 12.9 & 31 -- 36 \& 41 -- 45\\
\noalign{\smallskip}
2 & 0760646301 & 2015-09-02 & T & 15.9 & 10.2 & 31 -- 37 \& 39 -- 45\\
\noalign{\smallskip}
3 & 0760646401 & 2015-09-07 & T & 20.4 & 18.1 & 31 -- 36 \& 39 -- 45\\
\noalign{\smallskip}
4 & 0760646501 & 2015-09-12 & T & 18.9 & 18.9 & 31 -- 36 \& 40 -- 45\\
\noalign{\smallskip}
5 & 0760646601  & 2015-09-17 & SW & 52.1 & 11.8 & 26760.5, 27824.5,120,480\\
\noalign{\smallskip}
6 & 0760646701  & 2015-09-30 & SW & 47.7 & 36.3 & 26608.5, 27847.5,120,480\\
\noalign{\smallskip}
\hline\noalign{\smallskip} 
\end{tabular} 
\end{center}
Notes: \\
$^{\ddagger}$: T for timing mode, SW for small window mode\\ 
$^{\dagger}$: longest interval of continuous exposure\\
$^{*}$: for observations taken in timing mode: detector columns from which source photons have been selected; for observations taken in small window mode: x and y coordinate and inner and outer radius of the annulus region from which source photons have been selected in detector coordinates 
\label{Tab:Obs_xmm}
\end{table*}

\section[]{Observation and data analysis}
\label{Sec:obs}
\subsection{\swift\ monitoring}
The 2014/15 outburst of \gx339\ was detected by \swift/BAT \citep{2014ATel.6649....1Y} and followed-up with the \swift\ satellite \citep{2004ApJ...611.1005G,2005SSRv..120..143B}. We analysed all \swift/XRT \citep{2005SSRv..120..165B} observations taken in windowed timing mode between 2014 October 31st and 2015 October 3rd, using the online data analysis tools provided by the Leicester \swift\ data centre\footnote{http://www.swift.ac.uk/user\_objects/}, including single pixel events only \citep{2009MNRAS.397.1177E}.  

\subsection{\xmm}
\xmm\ \citep{2001A&A...365L...1J} followed the decay of the outburst with six observations. The first five observations are taken with a spacing of 5 days, while the sixth observation is taken 13 days after the fifth. Details on individual observations are given in Table~\ref{Tab:Obs_xmm}. We solely employed EPIC/pn data in our study, as they provide higher time resolution and are less affected by pile-up than data of the MOS cameras. We filtered and extracted the EPIC/pn event files using standard SAS (version 14.0.0) tools, paying particular attention to extract the list of photons not randomised in time. We analysed the data with the SAS task \texttt{epatplot} to check if the data are affected by pile-up and ended up using a source region that gives us an observed pattern distribution that follows the theoretical prediction quite nicely. We included single and double events (PATTERN$\le$4) in our study. 

For our timing studies we selected the longest interval of continuous exposure available in each observation. We produced PDS in the 1 -- 2  and 2 -- 10 keV band. We subtracted the contribution due to Poissonian noise \citep{1995ApJ...449..930Z}, normalised the PDS according to \citet{1983ApJ...272..256L} and converted to square fractional rms \citep{1990A&A...227L..33B}. 

To extract energy spectra we used the total exposures. We extracted energy spectra and corresponding background spectra, redistribution matrices, and ancillary response files for all observations.  

\subsection{\nus}
\nus\ \citep{2013ApJ...770..103H} observations taken simultaneously to the \xmm\ observations are available. Details on individual observations are given in Table~\ref{Tab:Obs_nus}. We analysed \nus\ data using the NuSTARDAS (version 1.4) tools \texttt{nupipeline} and \texttt{nuproducts}, with CALDB\,20160325. We extracted source photons from a circular region with a radius of 30'' located at the known position of \gx339. A background region of the same shape and size located close to the source on the same detector and free of source photons was used. To investigate short term variability we derived cospectra in the 3 -- 30 keV band using MaLTPyNT \citep{2015ascl.soft02021B}. The energy range between 3 and 30 keV comprises about 97 per cent of the source photons detected with \nus\ in the 3 -- 78 keV band. The cospectrum is the cross PDS derived from data of the two completely independent focal planes and represents a good proxy of the white-noise-subtracted PDS \citep{2015ApJ...800..109B}. 

\begin{table}
\caption{Details of \nus\ observations}
\begin{center}
\begin{tabular}{llrr}
\hline\noalign{\smallskip}
 \multicolumn{1}{c}{\#} & \multicolumn{1}{c}{Obs.~id.} & \multicolumn{1}{c}{Date}  &  \multicolumn{1}{c}{Exp. [ks]}  \\
 \hline\noalign{\smallskip}
1 & 80102011002  & 2015-08-28 & 21.6 \\
\noalign{\smallskip}
2 & 80102011004 & 2015-09-02 &  18.3 \\
\noalign{\smallskip}
3 & 80102011006 & 2015-09-07 &  19.8 \\
\noalign{\smallskip}
4 & 80102011008 & 2015-09-12 &  21.5 \\
\noalign{\smallskip}
5 & 80102011010  & 2015-09-17 &  38.5 \\
\noalign{\smallskip}
6 & 80102011012  & 2015-09-30 &  41.3 \\
\noalign{\smallskip}
\hline\noalign{\smallskip} 
\end{tabular} 
\end{center}
\label{Tab:Obs_nus}
\end{table}

\begin{figure}
\centering
\resizebox{\hsize}{!}{\includegraphics[clip,angle=0]{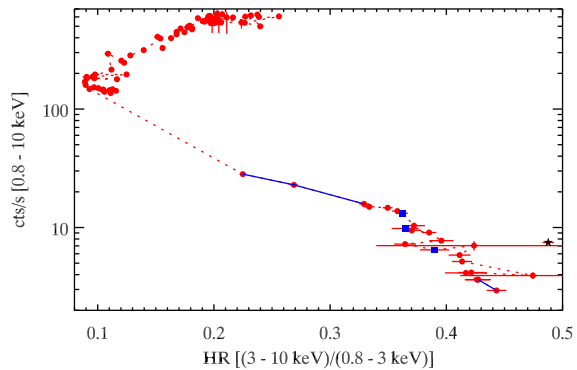}}
\caption{Hardness-intensity diagram, derived using \swift/XRT count rates. Each data point represents one observation. The first observation is marked by a star. The (blue) lines and squares indicate around which \swift/XRT observations the \xmm/\nus\ observations were taken.}
\label{Fig:HID}
\end{figure}

\begin{figure}
\centering
\resizebox{\hsize}{!}{\includegraphics[clip,angle=0]{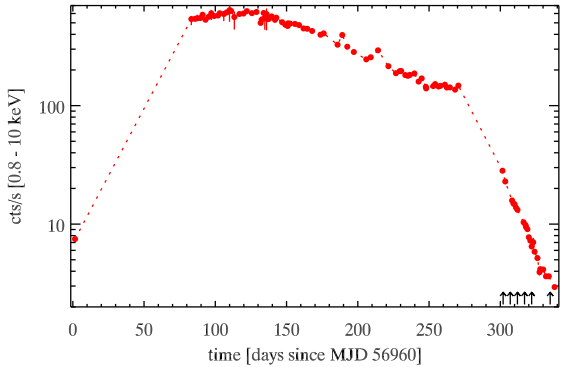}}
\caption{Light curve of the 2014/15 outburst, based on \swift/XRT count rates. Each data point represents one observation. The arrows mark the dates of the \xmm/\nus\ observations. T=0 corresponds to October 30th 2014 00:00:00.000 UTC.}
\label{Fig:LC}
\end{figure}

\section[]{Results}
\label{Sec:res}

\subsection{Hardness intensity diagram and light curve}
Based on the data available from the \swift/XRT monitoring of the outburst, we determined the source count rates in the total (0.8 -- 10 keV), soft (0.8 -- 3 keV), and hard (3 -- 10 keV) energy band, and derived a hardness ratio by dividing the count rate observed in the hard band by the one obtained in the soft band. The HID and the long term light curve is shown in Fig.~\ref{Fig:HID}  and Fig.~\ref{Fig:LC}, respectively. Due to sun constraints there is a gap of 81 days between the first (2014 Oct.\ 31) and second (2015 Jan.\ 21) \swift/XRT observation, where \gx339\ was already on its way to the soft state. The source entered into the HSS and stayed there at least until 2015 July 27. In the next available observation, taken after a gap of 31days, \gx339\ is already in the intermediate states on its way back to the LHS. In the light curve the times when the \xmm/\nus\ observations took place are indicated. 

\begin{figure}
\centering
\resizebox{\hsize}{!}{\includegraphics[clip,angle=0]{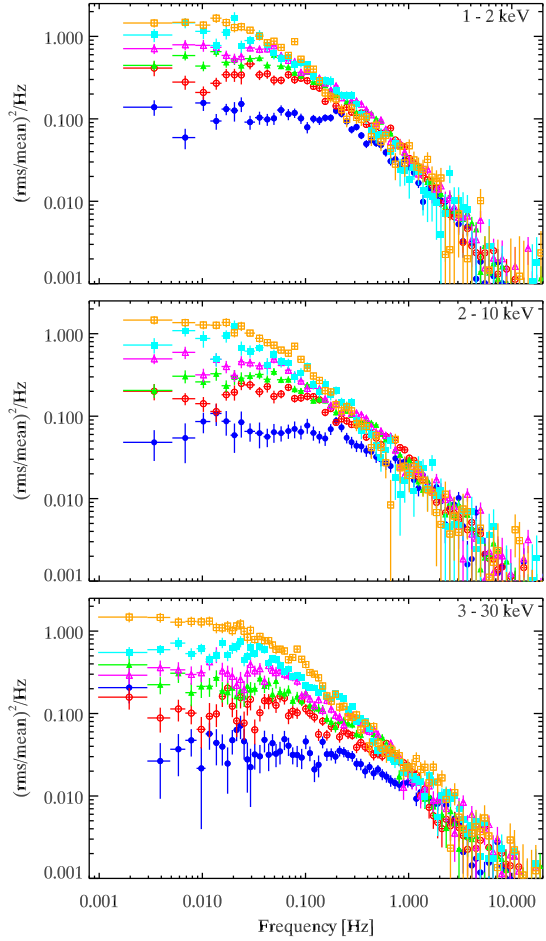}}
\caption{Power density spectra in the 1 -- 2 (upper panel), 2 -- 10 keV (middle panel), and 3 -- 30 keV (lower panel) bands. Different symbols indicate different observations: filled circles: obs.\ 1, open circles: obs.\ 2, filled triangles: obs.\ 3, open triangles: obs.\ 4, filled squares: obs.\ 5, open squares: obs.\ 6.}
\label{Fig:pds}
\end{figure}

\subsection{Power density spectra}
\begin{figure}
\centering
\resizebox{\hsize}{!}{\includegraphics[clip,angle=0]{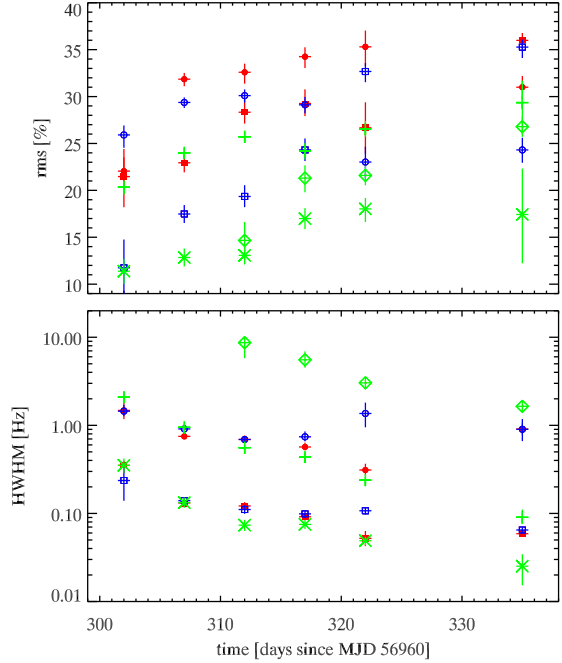}}
\caption{Evolution of the rms variability (upper panel) and characteristic frequency (lower panel). For \xmm\ observations, filled (red) symbols denote the soft band, while open (blue) symbols denote the 2 -- 10 keV band. Parameters of BLN1 are denoted by circles, those of BLN2 by squares. For \nus\ data, parameters of BNL1 are marked by `+' signs, of BLN2 by `x' signs, and of BLN3 by diamond shapes.}
\label{Fig:pdsprop}
\end{figure}

For the \xmm\ data, we derived PDS in the 1 -- 2 and 2 -- 10 keV range, for the \nus\ data we derived cospectra in the 3 -- 30 keV range (Fig.~\ref{Fig:pds}). For \xmm\ data, we used time bins of three times the frame time, which allowed us to sample frequencies up to $\sim28-29$ Hz, and stretches of 16384 bins. For \nus\ data we used time bins of $2^{-8}$ s and stretches of 512 s.
In general the PDS show band limited noise (BLN) components, fitted with zero-centered Lorentzians. Parameters can be found in Table~\ref{Tab:PDS}. For the \xmm\ data the soft and broadband PDS can be fitted with two BLN components. The same is true for the first two \nus\ observations, while the remaining four \nus\ observations require three BLN components. The additional BLN component shows up at a rather high characteristic frequency ($\sim$8.7 Hz) that decreases with ongoing outburst decay. An overall decrease of the characteristic frequency also shows up in the other two BLN components of the \nus\ data. In the \xmm\ PDS the decreasing trend in the characteristic frequency is less obvious (Fig.~\ref{Fig:pdsprop}). A QPO at a characteristic frequency of $\nu_{\mr{char}}=\sqrt{\nu_0^2+\Delta^2}\sim0.08 - 0.09$Hz is present in obs.\ 6. Parameters, including peak frequency ($\nu_0$) and half width at half maximum ($\Delta$), are given in Table~\ref{Tab:PDS}. Its significance decreases with higher energy ($3.7\sigma$ in the 1 -- 2 keV band, $2.8\sigma$ in the 2 -- 10 keV band, and $2.4\sigma$ in the 3 -- 30 keV band) and its Q factor is high ($Q=\nu_0/2\Delta\ga20$). The feature is a type-C QPO \citep{2005ApJ...629..403C,2011BASI...39..409B}. No harmonics are detected, which is most likely related to the fact that the observation is taken late during outburst decay when the source is already rather faint. In the \xmm\ data of obs.\ 1 a peaked noise component  at $\nu_{\mr{char}}\sim0.21$Hz is present (parameters are given in Table~\ref{Tab:PDS}). Its significance in the 1 -- 2 keV band ($2.8\sigma$) is higher than in the 2 -- 10 keV band ($2.3\sigma$), while the Q factor in the soft band (4.0) is smaller than in the 2 -- 10 keV band (5.3). 

\begin{figure}
\centering
\resizebox{\hsize}{!}{\includegraphics[clip,angle=0]{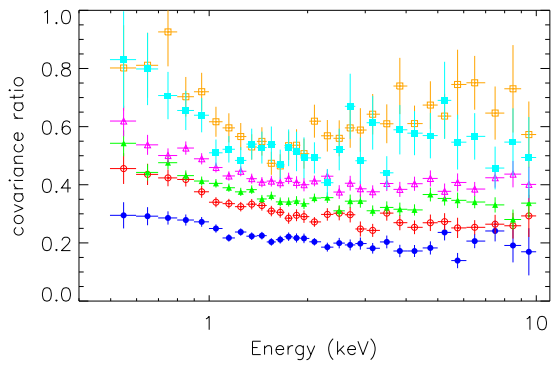}}
\caption{Covariance ratios for all six observations derived dividing the long timescale covariance spectra by the short timescale ones. Different (colours and) symbols indicate different observations: (blue) filled circles: obs.\ 1, (red) open circles: obs.\ 2, (green) filled triangles: obs.\ 3, (magenta) open triangles: obs.\ 4, (cyan) filled squares: obs.\ 5, (orange) open squares: obs.\ 6.}
\label{Fig:covr}
\end{figure}

\begin{table}
\caption{Parameters of the BLN components of the PDS}
\begin{center}
\footnotesize
\begin{tabular}{lrrr}
\hline\noalign{\smallskip}
 \multicolumn{1}{c}{Parameter} & \multicolumn{1}{c}{1 -- 2 keV} & \multicolumn{1}{c}{2 -- 10 keV} & \multicolumn{1}{c}{3 -- 30 keV} \\
\hline\noalign{\smallskip}
\multicolumn{4}{c}{Observation 1}\\
\hline\noalign{\smallskip}
rms$_{\mr{BLN1}}$ [\%] & $22.1^{+2.4}_{-2.0}$ & $25.9^{+1.0}_{-1.4}$ & $20.3^{+0.7}_{-0.8}$\\
\smallskip
$\nu_{\mr{BLN1}}$ [Hz] & $1.48^{+0.34}_{-0.30}$ & $1.45^{+0.26}_{-0.18}$ & $2.09^{+0.29}_{-0.26}$\\
\smallskip
rms$_{\mr{BLN2}}$ [\%] & $21.5^{+2.0}_{-3.3}$ & $11.7^{+3.0}_{-3.2}$& $11.4^{+1.3}_{-1.4}$ \\
\smallskip
$\nu_{\mr{BLN2}}$ [Hz] & $0.35^{+0.06}_{-0.05}$ & $0.24\pm0.10$ & $0.35^{+0.07}_{-0.06}$\\
\smallskip
rms$_{\mr{PN}}$ [\%] & $6.1^{+1.2}_{-1.1}$ & $5.1^{+1.4}_{-1.1}$& -- \\
\smallskip
$\nu_{\mr{0;PN}}$ [Hz] & $0.209^{+0.008}_{-0.007}$ & $0.205^{+0.006}_{-0.007}$ & -- \\
\smallskip
$\Delta_{\mr{PN}}$ [Hz] & $0.026^{+0.016}_{-0.012}$ & $0.019^{+0.016}_{-0.013}$ & --\\
\hline\noalign{\smallskip}
\multicolumn{4}{c}{Observation 2}\\
\hline\noalign{\smallskip}
rms$_{\mr{BLN1}}$ [\%] & $31.9^{+0.7}_{-0.8}$ & $29.4^{+0.5}_{-0.6}$ & $24.0^{+0.5}_{-0.6}$\\
\smallskip
$\nu_{\mr{BLN1}}$ [Hz] & $0.75^{+0.43}_{-0.37}$ & $0.91^{+0.06}_{-0.05}$ & $0.95^{+0.08}_{-0.07}$\\
\smallskip
rms$_{\mr{BLN2}}$ [\%] & $22.9^{+1.1}_{-1.0}$ & $17.5\pm1.0$& $12.8\pm1.0$ \\
\smallskip
$\nu_{\mr{BLN2}}$ [Hz] & $0.13\pm0.01$ & $0.14\pm0.01$ & $0.13\pm0.02$\\
\hline\noalign{\smallskip}
\multicolumn{4}{c}{Observation 3}\\
\hline\noalign{\smallskip}
rms$_{\mr{BLN1}}$ [\%] & $32.6^{+0.9}_{-1.2}$ & $30.1^{+0.6}_{-0.7}$ & $25.7\pm0.6$\\
\smallskip
$\nu_{\mr{BLN1}}$ [Hz] & $0.69^{+0.08}_{-0.05}$ & $0.69^{+0.06}_{-0.05}$ & $0.56^{+0.05}_{-0.04}$\\
\smallskip
rms$_{\mr{BLN2}}$ [\%] & $28.3^{+1.5}_{-1.2}$ & $19.4^{+1.2}_{-1.1}$& $13.1^{+0.9}_{-1.0}$ \\
\smallskip
$\nu_{\mr{BLN2}}$ [Hz] & $0.12\pm0.01$ & $0.11\pm0.01$ & $0.07\pm0.01$\\
\smallskip
rms$_{\mr{BLN3}}$ [\%] & -- & -- & $14.7^{+2.0}_{-2.1}$ \\
\smallskip
$\nu_{\mr{BLN3}}$ [Hz] &  -- & -- & $8.69^{+1.31}_{-2.88}$\\
\hline\noalign{\smallskip}
\multicolumn{4}{c}{Observation 4}\\
\hline\noalign{\smallskip}
rms$_{\mr{BLN1}}$ [\%] & $34.2^{+1.0}_{-1.2}$ & $29.1^{+0.8}_{-0.9}$ & $24.3^{+0.8}_{-0.9}$\\
\smallskip
$\nu_{\mr{BLN1}}$ [Hz] & $0.57^{+0.07}_{-0.05}$ & $0.74^{+0.11}_{-0.08}$ & $0.44\pm0.05$\\
\smallskip
rms$_{\mr{BLN2}}$ [\%] & $29.3^{+1.5}_{-1.3}$ & $24.3\pm1.2$& $17.1^{+1.0}_{-1.2}$ \\
\smallskip
$\nu_{\mr{BLN2}}$ [Hz] & $0.09\pm0.01$ & $0.10\pm0.01$ & $0.07\pm0.01$\\
\smallskip
rms$_{\mr{BLN3}}$ [\%] & -- & -- & $21.3^{+1.4}_{-1.5}$ \\
\smallskip
$\nu_{\mr{BLN3}}$ [Hz] &  -- & -- & $5.65^{+1.27}_{-1.13}$\\
\hline\noalign{\smallskip}
\multicolumn{4}{c}{Observation 5}\\
\hline\noalign{\smallskip}
rms$_{\mr{BLN1}}$ [\%] & $35.3^{+1.7}_{-2.1}$ & $23.0^{+1.6}_{-1.8}$ & $26.5^{+0.9}_{-0.8}$\\
\smallskip
$\nu_{\mr{BLN1}}$ [Hz] & $0.31^{+0.06}_{-0.04}$ & $1.36^{+0.45}_{-0.41}$ & $0.24\pm0.02$\\
\smallskip
rms$_{\mr{BLN2}}$ [\%] & $26.7\pm2.7$ & $32.7^{+0.9}_{-1.1}$& $18.0^{+1.1}_{-1.4}$ \\
\smallskip
$\nu_{\mr{BLN2}}$ [Hz] & $0.05\pm0.01$ & $0.11\pm0.01$ & $0.05\pm0.01$\\
\smallskip
rms$_{\mr{BLN3}}$ [\%] & -- & -- & $21.6^{+1.0}_{-1.1}$ \\
\smallskip
$\nu_{\mr{BLN3}}$ [Hz] &  -- & -- & $3.04^{+0.51}_{-0.41}$\\
\hline\noalign{\smallskip}
\multicolumn{4}{c}{Observation 6}\\
\hline\noalign{\smallskip}
rms$_{\mr{BLN1}}$ [\%] & $31.0^{+1.2}_{-1.3}$ & $24.3^{+1.3}_{-1.4}$ & $29.4^{+2.3}_{-3.4}$\\
\smallskip
$\nu_{\mr{BLN1}}$ [Hz] & $0.91^{+0.14}_{-0.13}$ & $0.90^{+0.28}_{-0.24}$ & $0.09^{+0.02}_{-0.01}$\\
\smallskip
rms$_{\mr{BLN2}}$ [\%] & $36.0^{+0.8}_{-0.9}$ & $35.3^{+0.8}_{-1.2}$& $17.4^{+4.9}_{-5.2}$ \\
\smallskip
$\nu_{\mr{BLN2}}$ [Hz] & $0.059\pm0.004$ & $0.065^{+0.004}_{-0.005}$ & $0.03\pm0.01$\\
\smallskip
rms$_{\mr{BLN3}}$ [\%] & -- & -- & $26.8\pm1.1$ \\
\smallskip
$\nu_{\mr{BLN3}}$ [Hz] &  -- & -- & $1.65\pm0.24$\\
\smallskip
rms$_{\mr{QPO}}$ [\%] & $6.8^{+1.2}_{-0.9}$ & $5.5^{+1.5}_{-1.0}$& $4.4\pm0.94$ \\
\smallskip
$\nu_{\mr{0;QPO}}$ [Hz] & $0.087^{+0.001}_{-0.003}$ & $0.088^{+0.001}_{-0.004}$ & $0.083\pm0.001$ \\
\smallskip
$\Delta_{\mr{QPO}}$ [Hz] & $<0.003$ & $<0.005$ & $0.002^{+0.002}_{-0.001}$\\
\hline\noalign{\smallskip} 
\end{tabular} 
\end{center}
\label{Tab:PDS}
\end{table}

\subsection{Covariance spectra}
\begin{figure*}
\centering
\resizebox{\hsize}{!}{\includegraphics[clip,angle=0]{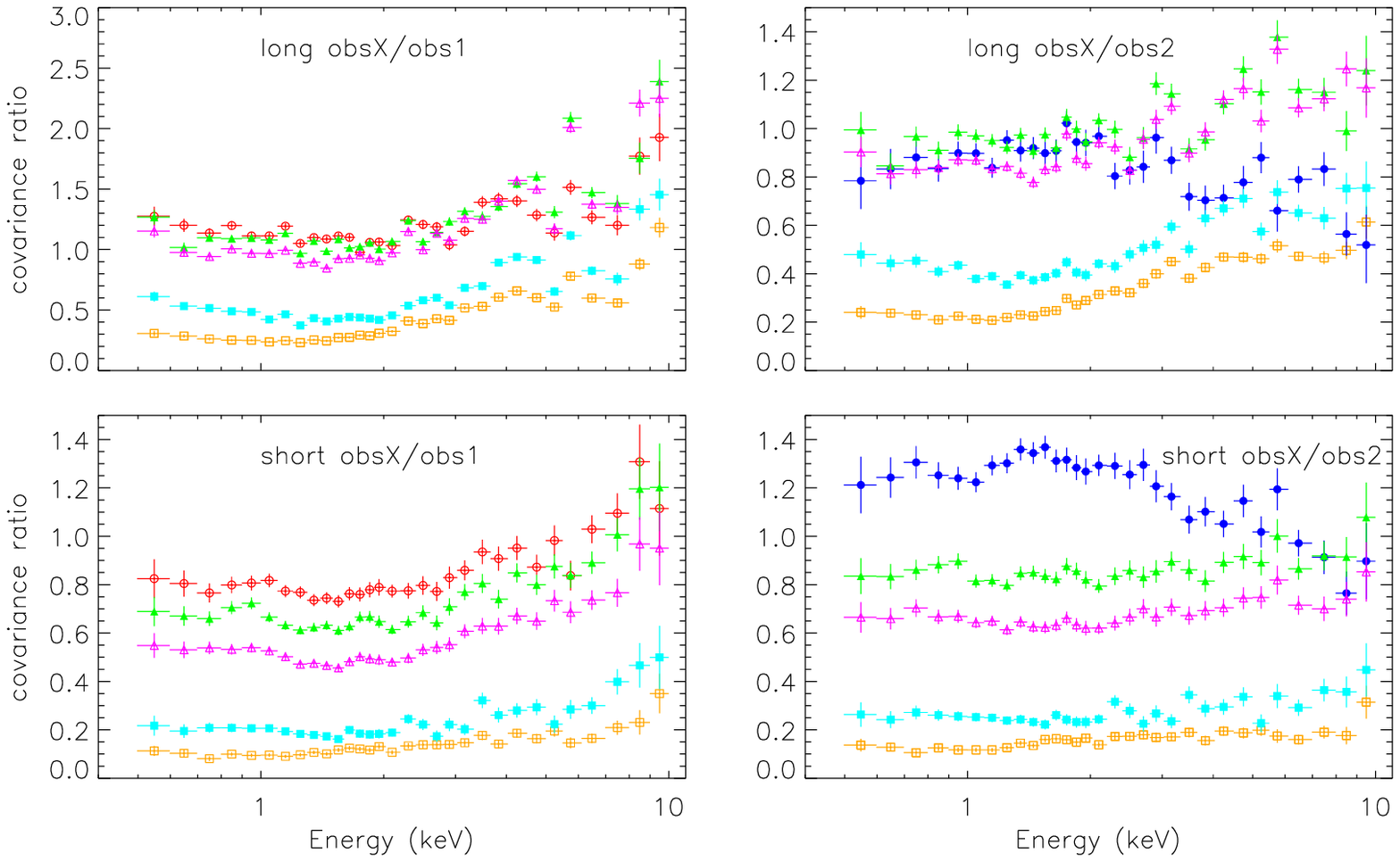}}
\caption{Covariance ratios for all observations derived dividing the short (lower panels) and long (upper panels) timescale covariance spectra by the corresponding spectra of obs.\ 1 (left panels) and obs.\ 2 (right panels). Symbols (and colours) as in Fig.\ \ref{Fig:covr}.}
\label{Fig:covr2}
\end{figure*}

Using \xmm\ data, which cover energies below 3 keV, we derived covariance spectra on short and long timescales and obtained covariance ratios dividing the long timescale covariance spectrum by the short timescale one \citep{2009MNRAS.397..666W,2015MNRAS.452.3666S}. We used the energy range between 1 and 4 keV as reference band, taking care to exclude energies from the reference band that are in the channel of interest. Taking a look at the PDS, we found that the break frequency where the top-flat part of the PDS goes into the decaying part evolves between different observations (see Fig.~\ref{Fig:pds}). We can use the following timescales for all observations to compare variability in the decaying part of the PDS to variability in the flat part: 0.05~s time bins measured in segments of 3.5~s for shorter time scales and 12.5~s time bins measured in segments of 625~s for longer time scales. The obtained covariance ratios are shown in Fig.~\ref{Fig:covr}, where different symbols indicate different observations. We find that the (energy-averaged) covariance ratio increases during outburst decay. Furthermore, we find in all observations at energies below $\sim 2$ keV an increase of the covariance ratios towards lower energies. This behaviour has also been observed for observations of \gx339\ taken during outburst rise and has been interpreted as sign of additional disk variability on longer timescales \citep{2009MNRAS.397..666W,2015MNRAS.452.3666S}. Comparing the long-to-short covariance ratios of different observations with each other reveals that the increase of the covariance ratio steepens with time. While for the first four observations the covariance ratios above 2 keV remain rather flat, we find an increase towards higher energies in the last two observations. 
In addition to study covariance ratios by comparing spectra on long and short timescales for each observation, we can derive ratios by comparing variability spectra on short (or long) timescales between different observations (Fig.~\ref{Fig:covr2}). The ratios obtained by dividing the short and long timescale covariance spectra of all observations by those of the first observation, show an increase towards higher energies. This indicates that \gx339\ hardens during outburst decay. For the short timescale ratios, we find that the ratio decreases during outburst decay, which is to be expected as the source gets fainter during outburst decay. On the long timescale \gx339\ first gets brighter before it reaches again in obs.\ 4 a brightness comparable to that of the first observation, and then gets fainter in the last two observations. Deriving ratios relative to obs.\ 2, we see that on short timescales the ratios (of the later observations) are rather flat, indicating that not much spectral evolution takes place on short timescales after obs.\ 2, while the spectral hardening continues on the longer timescales.

\subsection{Energy spectra}
We fit averaged energy spectra, using simultaneous \xmm/EPICpn and \nus\ data, within \textsc{isis} \citep[V.~1.6.2;][]{2000ASPC..216..591H} in the 0.8 -- 78 keV range, where \xmm\ data covered the 0.8 -- 10 keV range and \nus\ data the 4 -- 78 keV range. We excluded \nus\ data between 3 and 4 keV from our analysis, as the spectral residuals show a clear mismatch of NuSTAR in this energy range. This behaviour has been observed previously and is attributed to pile-up and cross-calibration differences \citep{2016ApJ...819..150F}, which are taken care of in newer versions of NuSTARDAS and CALDB.
We grouped the \xmm\ data to a minimum signal to noise ratio of three and the \nus\ data to a minimum signal to noise ratio of five. For the four \xmm\ observations taken in timing mode, we ignored energies between 2.0 and 2.4 keV, as this energy range is affected by a residual feature related to small shifts in energy gain at the Si-K and Au-M edges of the instrumental response \citep{2011MNRAS.416..311K}. To fit the energy spectra we used the \texttt{relxill} model \citep{2014MNRAS.444L.100D,2014ApJ...782...76G} together with a disk blackbody component \citep[\texttt{diskbb};][]{1984PASJ...36..741M}. The \texttt{relxill} model allows us to fit the Comptonized emission including relativistic reflection. We included an absorption component \citep[\texttt{TBabs};][]{2000ApJ...542..914W}, using the abundances of \citet{2000ApJ...542..914W} and the cross sections given in \citet{1996ApJ...465..487V}. We also added a floating cross-normalization parameter, which was fixed to one for \nus\ FPMA, to take uncertainties in the cross-calibration between the different telescopes into account. 
We constrained the foreground absorption to \nh$=5.55\pm0.09$\hcm{21}, fixed the inclination at 30\deg\ and the spin at 0.95 \citep{2016ApJ...821L...6P}. We used an emissivity index of 3 \citep{1997ApJ...488..109R}. \citet{2015ApJ...808..122F} showed that allowing for a variable emissivity index all other parameters did not change significantly.  

Allowing for a free ionization parameter and iron abundance in the \texttt{relxill} model, we find that these two parameters show large variability between different observations. The ionization parameter in obs.\ 2, 3 and 4 is close to zero, indicating a neutral disk, while the disk seems to be highly ionised in obs.\ 1 and 5, and slightly ionised in obs.\ 6. The iron abundance decreases from $\sim$6 in the first observation down to $\sim$2 in obs.\ 3 and then increases to the maximum allowed value of 10 in the last observation. This behaviour, especially the one of the iron abundance, seems to be highly unphysical. \citet{2015ApJ...808..122F} addressed a problem with a too high iron abundance by allowing the photon-indices of the continuum and the input to the reflector to be different. Including an additional \texttt{nthcomp} \citep{1996MNRAS.283..193Z,1999MNRAS.309..561Z} component, we find an unphysical high iron abundance of 10 in all observations, and an ionization parameter of 3 -- 4 in all observations but the last one, where it is close to zero. \citet{2016MNRAS.458.2199B}, who reanalysed seven archival \xmm\ observations of \gx339\ in the hard state at outburst rise with a model similar to the one used in our study, also found that the iron abundance pegged at the maximum allowed value when the \texttt{diskbb} normalisation is a free parameter. Since adding an \texttt{nthcomp} component does not help to get a more physical iron abundance, we fixed $A_{\mr{Fe}}$ at the best fit value of 1.58 obtained by \citet{2015ApJ...808..122F}. As we still find a very low ionization parameter in obs.\ 2, 3, and 4 and a high ionization parameter in obs.\ 1, 5, and 6, we decide to fit all observations either with a low or a high ionization parameter.

\begin{figure*}
\centering
\resizebox{\hsize}{!}{\includegraphics[clip,angle=0]{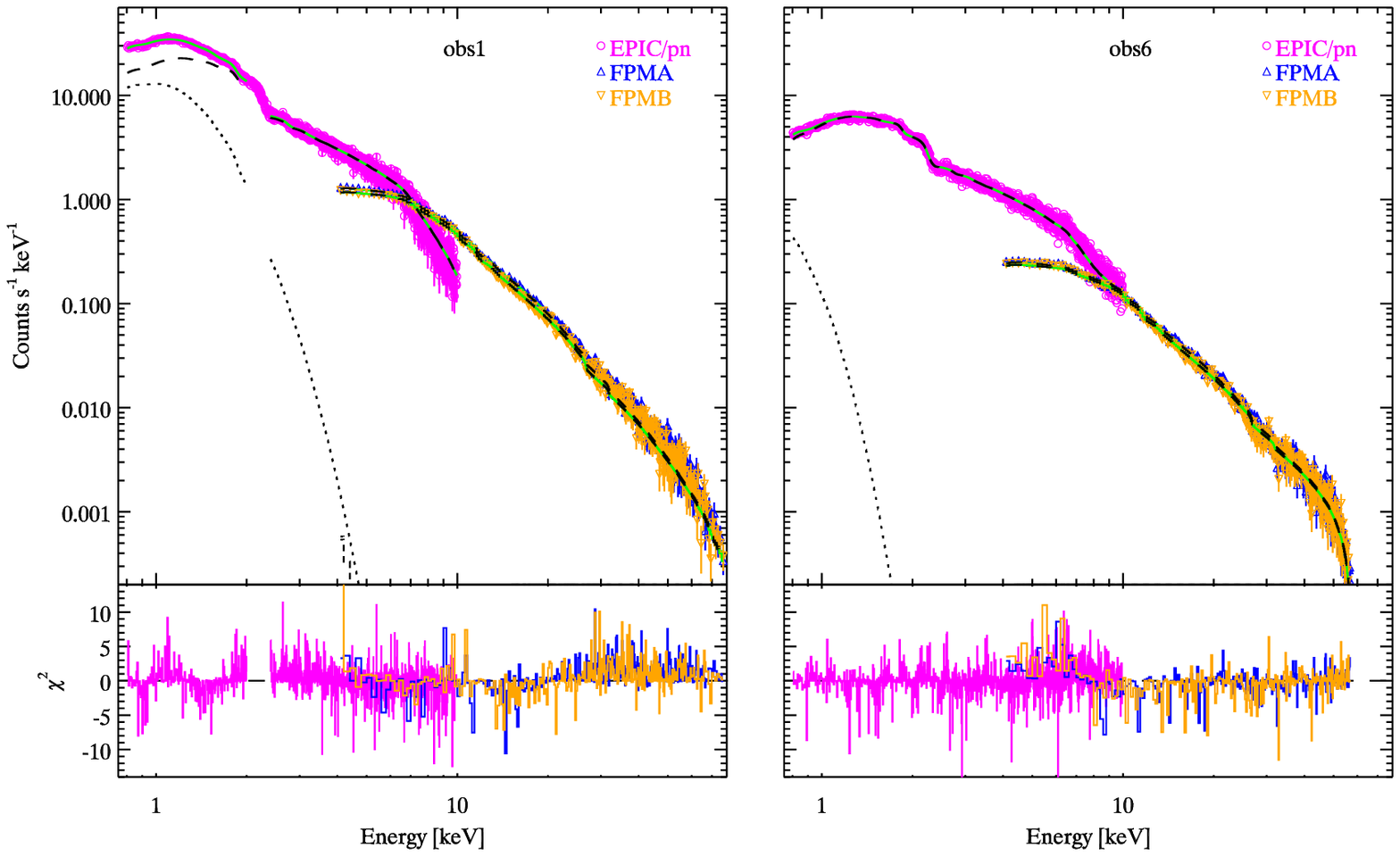}}
\caption{Combined \xmm/EPICpn and \nus\ spectra of Obs.\ 1 (left panel) and Obs.\ 6 (right panel). Data points, best fit model (see Table~\ref{Tab:specpar}) and individual model components are given (disk blackbody: dotted line; \texttt{relxill}: dashed line). For Obs.\ 1 the best fit model assuming a highly ionised reflection disk is shown, while for Obs.\ 6 the model with a neutral reflection disk is shown.}
\label{Fig:espec}
\end{figure*}

\begin{figure*}
\centering
\resizebox{\hsize}{!}{\includegraphics[clip,angle=0]{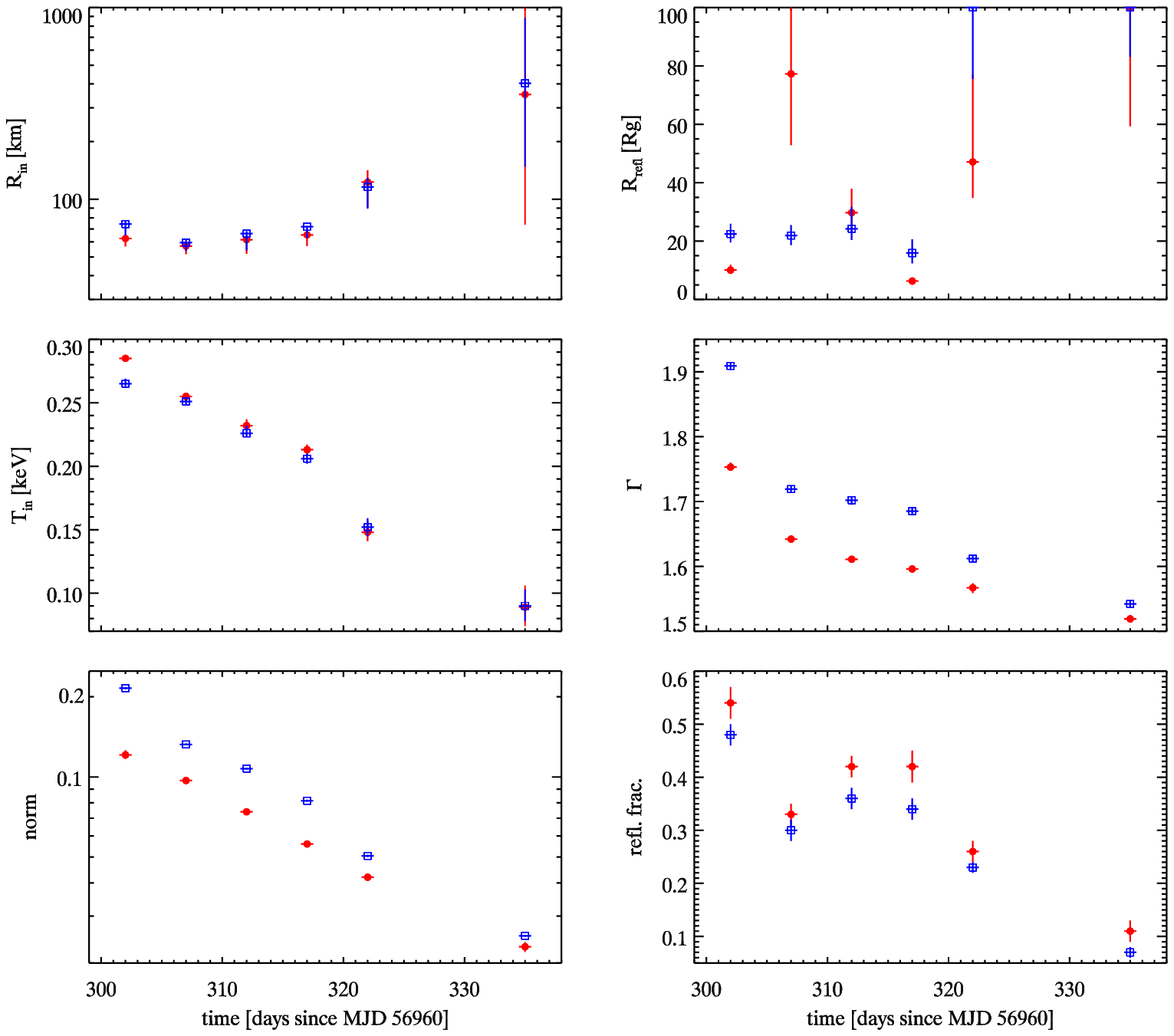}}
\caption{Evolution of the inner disk radius (from \texttt{diskbb},assuming a distance to \gx339\ of 8.4 kpc and an inclination of 30\deg \citep{2016ApJ...821L...6P}), disk temperature, \texttt{relxill} norm, inner disk radius (from \texttt{relxill}), photon index, and reflection fraction.}
\label{Fig:specpar}
\end{figure*}

The energy spectra of the first and last observation are shown in Fig.~\ref{Fig:espec}. The obtained spectral parameters are given in table~\ref{Tab:specpar} and how they evolve during outburst decay is shown in Fig.~\ref{Fig:specpar}. We find that the inner disk radius obtained from the \texttt{diskbb} model increases, while the disk temperature, photon index and \texttt{relxill} normalization decrease monotonically along outburst decay. We derive the inner disk radius from the normalisation of the \texttt{diskbb} model, assuming a distance to \gx339\ of $8.4\pm0.9$ kpc and an inclination of $30$\deg$\pm1$\deg \citep{2016ApJ...821L...6P}, and including a correction factor of 1.18, to account for spectral hardening (assuming a hardening factor of 1.7) and for the fact that the disk temperature does not peak at the inner radius \citep{2013ApJ...769...16R}. The reflection fraction, which is always below unity, also shows a decreasing trend. Small reflection fractions below 0.5 have also been measured for the LHS at outburst rise and can indicate either a truncated accretion disk or an outflowing corona \citep{2015ApJ...808..122F}. Assuming a neutral reflection disk the radius of the reflector obtained from the \texttt{relxill} model is $20\pm5\ R_g$ and then increases to more than 75  $R_g$ in the last two observations. In the case of a highly ionised disk the radius of the reflector is highly variable ($5.5<R_{\mr{refl}}<100\ R_g$). For the last two observations the radius of the reflector of a highly ionised disk is consistent within errors with the radii for a neutral reflection disk.

Assuming the lamppost geometry (\texttt{relxilllp}), and allowing for a free iron abundance, ionisation parameter, reflection disk radius and lamppost height, we obtain an iron abundance around 5, which increases to the maximum allowed value of 10 in the last two observations, an ionisation parameter indicating a highly ionized disk, a reflection disk radius below $\sim3\ R_g$, apart from the last observation where it is not constrained, and a highly variable lamppost height between $\sim 4$ and 25 $R_g$, apart from the last observation where it is not constrained. Assuming an iron abundance of 1.58 and a highly ionised disk, we obtain an increasing reflection disk radius, apart from obs.\ 4, and a variable lamppost height, which is not constrained in the last two observations. The spectral parameters are given in table~\ref{Tab:specpar_lp}. Fixing the reflection disk radius at the ISCO, does not help to reduce the variability in the lamppost height.

\begin{table*}
\caption{Parameters of the simultaneous \xmm/EPICpn and \nus\ energy spectra}
\begin{center}
\begin{tabular}{lrrrrrr}
\hline\noalign{\smallskip}
 \multicolumn{1}{c}{Parameter} & \multicolumn{1}{c}{Obs.\ 1} & \multicolumn{1}{c}{Obs.\ 2} & \multicolumn{1}{c}{Obs.\ 3} & \multicolumn{1}{c}{Obs.\ 4} & \multicolumn{1}{c}{Obs.\ 5} & \multicolumn{1}{c}{Obs.\ 6}\\
\hline\noalign{\smallskip}
\multicolumn{7}{c}{$A_{\mr{Fe}}=1.58$; reflection disk highly ionized}\\
\hline\noalign{\smallskip}
$R_{\mr{in}}^{\dagger}$ (km) & $ 62.41_{-11.78}^{+3.52}$ & $56.99_{-10.82}^{+3.26}$ & $61.54_{-15.09}^{+4.37}$ & $65.15_{-14.26}^{+6.32}$ & $122.88_{-42.12 }^{+34.19}$ & $352.14_{-280.51}^{+787.43} $ \\     
\smallskip
$T_{\mr{in}}$ (keV)      &$ 0.285_{-0.002}^{+0.002}$ & $  0.255_{-0.002}^{+0.002}$ & $ 0.232_{-0.003}^{+0.005}$ & $ 0.213_{-0.004}^{+0.004}$ & $ 0.148 _{-0.007}^{+0.009}$ & $ 0.089_{-0.015}^{+0.017 }$   \\    
\smallskip
norm$_{\mr{rx}}$     & $ 0.121_{-0.004}^{+0.005}$ & $  0.097_{-0.001}^{+0.001}$ & $ 0.074_{-0.001}^{+0.001}$ & $ 0.056_{-0.001}^{+0.001}$ & $ 0.042 _{-0.001}^{+0.001}$ & $ 0.023_{-0.001}^{+0.001 }$  \\
\smallskip
$R_{\mr{refl}}$ ($R_g$)     & $ 10.07_{-1.33 }^{+1.85 }$ & $  77.27_{-24.40}^{+22.73}$ & $ 29.74_{-6.65 }^{+8.19 }$ & $ 6.32 _{-0.78 }^{+0.99 }$ & $ 47.14 _{-12.36}^{+29.93}$ & $ 100.00_{-40.64}^{+0.00 }$ \\       
\smallskip
$\Gamma$    & $ 1.753_{-0.005}^{+0.007}$ & $  1.642_{-0.003}^{+0.003}$ & $ 1.611_{-0.003}^{+0.003}$ & $ 1.596_{-0.003}^{+0.003}$ & $ 1.567 _{-0.008}^{+0.007}$ & $ 1.519_{-0.005}^{+0.006 }$  \\     
\smallskip
$\mathcal{R}$ &  $ 0.54 _{-0.03 }^{+0.03 }$ & $  0.33 _{-0.02 }^{+0.02 }$ & $ 0.42 _{-0.02 }^{+0.02 }$ & $ 0.42 _{-0.03 }^{+0.03 }$ & $ 0.26  _{-0.02 }^{+0.02 }$ & $ 0.11_{-0.02 }^{+0.02 }$ \\
\smallskip
$CC_{\mr{FPMB}}$   &  $ 0.967_{-0.003}^{+0.003}$ & $  0.995_{-0.004}^{+0.004}$ & $ 0.988_{-0.004}^{+0.004}$ & $ 0.993_{-0.004}^{+0.004}$ & $ 1.016 _{-0.004}^{+0.004}$ & $ 1.008 _{-0.005}^{+0.005 }$\\    
\smallskip
$CC_{\mr{pn}}$  & $ 1.626_{-0.007}^{+0.007}$ & $  1.903_{-0.007}^{+0.007}$ & $ 1.758_{-0.007}^{+0.007}$ & $ 1.705_{-0.008}^{+0.008}$ & $ 0.914 _{-0.005}^{+0.005}$ & $ 0.948 _{-0.005}^{+0.005 }$  \\    
\smallskip
$\chi^2$/dof&  2666.2/2155              &    2821.5/2341               &   2564.8/2326               &   2568.94/2276              &   2582.9/2251               &   2575.9/2274 \\
F$_{\mr{abs}}^{\ddagger}$ & 13.39                    &    11.67                     &   8.92                      &   6.60                      &   2.37                      &   1.31\\
F$_{\mr{unabs}}^{\ddagger}$ & 28.73                    &    25.56                     &   19.55                     &   14.50                     &   5.21                      &   2.87\\
\hline\noalign{\smallskip} 
\multicolumn{7}{c}{$A_{\mr{Fe}}=1.58$; neutral reflection disk}\\
\hline\noalign{\smallskip}
$R_{\mr{in}}^{\dagger}$ (km) &  $74.21_{-16.31}^{+4.29}$ & $59.42_{-11.36}^{+3.46}$ & $66.24_{-17.93}^{+5.07}$ & $71.89_{-15.81}^{+7.27}$ & $115.76_{-35.71}^{+27.93}$ & $402.82_{-267.13}^{+579.28}$\\     
\smallskip
$T_{\mr{in}}$ (keV)      & $0.265 _{-0.002 }^{+0.004 }$ & $ 0.251 _{-0.002 }^{+0.002 }$ & $ 0.226 _{-0.003 }^{+0.006 }$ & $ 0.206 _{-0.004 }^{+0.004 }$ & $ 0.152 _{-0.007 }^{+0.007 }$ & $ 0.090 _{-0.012 }^{+0.013}$  \\    
\smallskip
norm$_{\mr{rx}}$     &  $0.2156_{-0.0019}^{+0.0019}$ & $ 0.1325_{-0.0010}^{+0.0011}$ & $ 0.1075_{-0.0011}^{+0.0009}$ & $ 0.0814_{-0.0008}^{+0.0008}$ & $ 0.0505_{-0.0005}^{+0.0005}$ & $ 0.0253_{-0.0002}^{+0.0002}$ \\
\smallskip
$R_{\mr{refl}}$ ($R_g$)     & $22.45 _{-2.88  }^{+3.45  }$ & $ 21.91 _{-3.29  }^{+3.53  }$ & $ 24.24 _{-3.80  }^{+7.47  }$ & $ 15.93 _{-3.54  }^{+4.69  }$ & $ 100.00_{-24.41 }^{+0.00  }$ & $ 100.00_{-16.77 }^{+0.00}$ \\       
\smallskip
$\Gamma$    &   $1.909 _{-0.006 }^{+0.006 }$ & $ 1.719 _{-0.005 }^{+0.005 }$ & $ 1.702 _{-0.007 }^{+0.005 }$ & $ 1.685 _{-0.006 }^{+0.006 }$ & $ 1.612 _{-0.005 }^{+0.005 }$ & $ 1.542 _{-0.004 }^{+0.004}$\\     
\smallskip
$\mathcal{R}$ & $ 0.48  _{-0.02  }^{+0.02  }$ & $ 0.30  _{-0.02  }^{+0.02  }$ & $ 0.36  _{-0.02  }^{+0.02  }$ & $ 0.34  _{-0.02  }^{+0.02  }$ & $ 0.23  _{-0.01  }^{+0.01  }$ & $ 0.07  _{-0.01  }^{+0.01}$ \\
\smallskip
$CC_{\mr{FPMB}}$   & $0.967 _{-0.003 }^{+0.003 }$ & $ 0.995 _{-0.004 }^{+0.004 }$ & $ 0.988 _{-0.004 }^{+0.004 }$ & $ 0.993 _{-0.004 }^{+0.004 }$ & $ 1.015 _{-0.004 }^{+0.004 }$ & $ 1.008 _{-0.005 }^{+0.005}$\\    
\smallskip
$CC_{\mr{pn}}$  &  $1.600 _{-0.007 }^{+0.007 }$ & $ 1.903 _{-0.007 }^{+0.007 }$ & $ 1.754 _{-0.007 }^{+0.007 }$ & $ 1.695 _{-0.008 }^{+0.008 }$ & $ 0.909 _{-0.005 }^{+0.005 }$ & $ 0.948 _{-0.005 }^{+0.005}$ \\    
\smallskip
$\chi^2$/dof&  2646.2/2155                &   2836.3/2340                 &   2514.9/2325                  &   2506.4/2275                  &   2591.7/2251                  &   2596.2/2274  \\
F$_{\mr{abs}}^{\ddagger}$ & 13.54                      &   11.71                       &   8.96                         &   6.62                         &   2.35                         &   1.30\\
F$_{\mr{unabs}}^{\ddagger}$ & 29.78                      &   25.63                       &   19.62                        &   14.55                        &   5.18                         &   2.86\\
\hline\noalign{\smallskip}
\end{tabular} 
\end{center}
Notes: \\
$^{\dagger}$: assuming a distance to \gx339\ of $8.4\pm0.9$ kpc and an inclination of $30$\deg$\pm1$\deg \citep{2016ApJ...821L...6P}\\
$^{\ddagger}$: X-ray flux in the 0.8 -- 78 keV band in unites of \oergcm{-9}
\label{Tab:specpar}
\end{table*}

\begin{table*}
\caption{Parameters of the simultaneous \xmm/EPICpn and \nus\ energy spectra assuming lamppost geometry}
\begin{center}
\begin{tabular}{lrrrrrr}
\hline\noalign{\smallskip}
 \multicolumn{1}{c}{Parameter} & \multicolumn{1}{c}{Obs.\ 1} & \multicolumn{1}{c}{Obs.\ 2} & \multicolumn{1}{c}{Obs.\ 3} & \multicolumn{1}{c}{Obs.\ 4} & \multicolumn{1}{c}{Obs.\ 5} & \multicolumn{1}{c}{Obs.\ 6}\\
\hline\noalign{\smallskip}
\multicolumn{7}{c}{$A_{\mr{Fe}}=1.58$; reflection disk highly ionized}\\
\hline\noalign{\smallskip}
$R_{\mr{in}}^{\dagger}$ (km) & $ 62.43_{-11.61}^{+3.55}$ & $ 58.61_{-11.02}^{+3.27}$ & $ 65.26_{-13.09}^{+4.64}$ & $ 70.36_{-15.91}^{+7.51}$ & $ 120.38_{-40.06}^{+34.52}$ & $ 317.36_{-247.66}^{+695.98} $ \\     
\smallskip
$T_{\mr{in}}$ (keV)      &$ 0.285_{-0.002}^{+0.002}$ & $ 0.254_{-0.002}^{+0.002}$ & $ 0.228_{-0.003}^{+0.003}$ & $ 0.209_{-0.004}^{+0.004}$ & $ 0.155_{-0.001}^{+0.001}$ & $ 0.089_{-0.010}^{+0.016 }$   \\    
\smallskip
norm$_{\mr{rx}}$     & $ 0.1209_{-0.0044}^{+0.0045 }$ & $ 0.1123_{-0.0025}^{+0.0017 }$ & $ 0.0873_{-0.0018}^{+0.0019 }$ & $ 0.0582_{-0.0009}^{+0.0010 }$ & $ 0.0397_{-0.0001}^{+0.0001 }$ & $ 0.0226_{-0.0005}^{+0.0007}$  \\
\smallskip
$R_{\mr{refl}}$ ($R_g$)     & $  5.57 _{-5.57 }^{+3.19}$ & $ 12.71_{-12.71}^{+12.04}$ & $ 29.17_{-8.58 }^{+5.00}$ & $ 6.64 _{-1.85 }^{+0.98}$ & $ 59.84_{-29.92}^{+20.08}$ & $ 100.00_{-63.53}^{+0.00}$ \\       
\smallskip
h ($R_g$)     & $  15.11_{-4.73 }^{+4.43}$ & $ 53.04_{-18.56}^{+31.49}$ & $ 3.02 _{-0.02 }^{+23.29}$ & $ 3.48 _{-0.48 }^{+5.14}$ & $ 3.00 _{-0.00 }^{+97.00}$ & $ 100.00_{-97.00}^{+0.00}$ \\       
\smallskip
$\Gamma$    & $  1.754_{-0.005}^{+0.007}$ & $ 1.665_{-0.005}^{+0.005}$ & $ 1.638_{-0.006}^{+0.004}$ & $ 1.607_{-0.003}^{+0.003}$ & $ 1.554_{-0.003}^{+0.003}$ & $ 1.519_{-0.005}^{+0.006}$  \\     
\smallskip
$\mathcal{R}$ &  $  0.548_{-0.024}^{+0.035}$ & $ 0.245_{-0.013}^{+0.017}$ & $ 0.313_{-0.016}^{+0.008}$ & $ 0.413_{-0.029}^{+0.029}$ & $ 0.281_{-0.002}^{+0.005}$ & $ 0.109_{-0.020}^{+0.020}$ \\
\smallskip
$CC_{\mr{FPMB}}$   &  $  0.967_{-0.003}^{+0.003}$ & $ 0.995_{-0.004}^{+0.004}$ & $ 0.988_{-0.004}^{+0.004}$ & $ 0.993_{-0.004}^{+0.004}$ & $ 1.015_{-0.003}^{+0.003}$ & $ 1.008_{-0.005}^{+0.005}$\\    
\smallskip
$CC_{\mr{pn}}$  & $  1.626_{-0.006}^{+0.007}$ & $ 1.909_{-0.007}^{+0.007}$ & $ 1.762_{-0.007}^{+0.007}$ & $ 1.702_{-0.008}^{+0.008}$ & $ 0.914_{-0.002}^{+0.002}$ & $ 0.948_{-0.003}^{+0.005}$  \\    
\smallskip
$\chi^2$/dof&  2664.3/2155 & 3113.9/2339 & 2697.1/2324 & 2561.7/2275 & 2587.4/2251 & 2575.9/2273 \\
F$_{\mr{abs}}^{\ddagger}$ & 13.40 & 11.63 & 8.93 & 6.62 & 2.36 & 1.31 \\
F$_{\mr{unabs}}^{\ddagger}$ & 28.74 & 25.58 & 19.63 & 14.56 & 5.20 & 2.87\\
\hline\noalign{\smallskip} 
\multicolumn{7}{c}{$A_{\mr{Fe}}=1.58$; reflection disk highly ionized; reflection disk radius fixed at ISCO}\\
\hline\noalign{\smallskip}
$R_{\mr{in}}^{\dagger}$ (km) &  $ 62.44_{-11.35}^{+3.49}$ & $ 58.63_{-11.00}^{+3.25}$ & $ 65.18_{-13.02}^{+4.18}$ & $ 67.47_{-14.90}^{+6.79}$ & $ 103.17_{-19.10}^{+5.21}$ & $ 353.26_{-262.76}^{+788.78 }$\\     
\smallskip
$T_{\mr{in}}$ (keV)      & $ 0.285_{-0.002}^{+0.002}$ & $ 0.254_{-0.002}^{+0.002}$ & $ 0.227_{-0.003}^{+0.003}$ & $ 0.204_{-0.005}^{+0.005}$ & $ 0.148_{-0.000}^{+0.008}$ & $ 0.091_{-0.015}^{+0.016}$  \\    
\smallskip
norm$_{\mr{rx}}$     &  $0.120_{-0.004}^{+0.005}$ & $ 0.112_{-0.003}^{+0.002}$ & $ 0.087_{-0.003}^{+0.003}$ & $ 0.064_{-0.001}^{+0.001}$ & $ 0.042_{-0.001}^{+0.001}$ & $ 0.022_{-0.001}^{+0.001}$ \\
\smallskip
h ($R_g$)     & $  17.09_{-2.38 }^{+3.25}$ & $ 61.23_{-18.39}^{+32.03}$ & $ 35.20_{-7.09 }^{+11.43}$ & $ 10.45_{-1.65 }^{+2.17}$ & $ 96.40_{-36.26}^{+3.60}$ & $ 100.00_{-29.14}^{+0.00}$ \\          
\smallskip
$\Gamma$    &   $ 1.754_{-0.007}^{+0.007}$ & $ 1.665_{-0.005}^{+0.005}$ & $ 1.640_{-0.006}^{+0.006}$ & $ 1.627_{-0.004}^{+0.004}$ & $ 1.566_{-0.007}^{+0.007}$ & $ 1.517_{-0.004}^{+0.006}$\\     
\smallskip
$\mathcal{R}$ & $  0.56 _{-0.03 }^{+0.04}$ & $ 0.25 _{-0.02 }^{+0.02}$ & $ 0.32 _{-0.02 }^{+0.02}$ & $ 0.37 _{-0.03 }^{+0.03}$ & $ 0.26 _{-0.02 }^{+0.02}$ & $ 0.11 _{-0.02 }^{+0.02}$ \\
\smallskip
$CC_{\mr{FPMB}}$   & $ 0.967_{-0.003}^{+0.003}$ & $ 0.995_{-0.004}^{+0.004}$ & $ 0.988_{-0.004}^{+0.004}$ & $ 0.993_{-0.004}^{+0.004}$ & $ 1.016_{-0.004}^{+0.004}$ & $ 1.008_{-0.005}^{+0.005}$\\    
\smallskip
$CC_{\mr{pn}}$  &  $ 1.625_{-0.007}^{+0.007}$ & $ 1.909_{-0.007}^{+0.007}$ & $ 1.762_{-0.007}^{+0.007}$ & $ 1.704_{-0.008}^{+0.008}$ & $ 0.914_{-0.005}^{+0.005}$ & $ 0.948_{-0.005}^{+0.005}$ \\    
\smallskip
$\chi^2$/dof&   2664.9/2155  & 3114.4/2340 & 2698.6/2325 & 2556.9/2276 & 2585.1/2251 & 2577.8/2274 \\
F$_{\mr{abs}}^{\ddagger}$ & 13.40 & 11.63 & 8.94 & 6.67 & 2.37 & 1.31\\
F$_{\mr{unabs}}^{\ddagger}$ & 28.75 & 25.57 & 19.66 & 14.68 & 5.21 & 2.87\\
\hline\noalign{\smallskip}
\end{tabular} 
\end{center}
Notes: \\
$^{\dagger}$: assuming a distance to \gx339\ of $8.4\pm0.9$ kpc and an inclination of $30$\deg$\pm1$\deg \citep{2016ApJ...821L...6P}\\
$^{\ddagger}$: X-ray flux in the 0.8 -- 78 keV band in unites of \oergcm{-9}
\label{Tab:specpar_lp}
\end{table*}

\section[]{Discussion}
\label{Sec:dis}
In this study we made use of simultaneous \xmm\ and \nus\ data obtained during the decay of the 2014/15 outburst of \gx339, to investigate the evolution of the energy spectra and variability during outburst decay in the 0.8 to 78 keV range. The HID derived from the available \swift/XRT monitoring data clearly shows that the \xmm/\nus\ observations of \gx339\ followed the evolution of the source 
in the decaying branch of the outburst. 

The coexisting of two different power spectral shapes in the soft and hard band during the hard-to-soft state transition has been reported perviously \citep{2013ApJ...770..135Y,2014MNRAS.441.1177S}. In this study, the shape of the power density spectra is the same in the 1 -- 2 and 2 -- 10 keV band, as the PDS of both bands can be fitted by two BLN components. Thus, there is no sign of energy dependence of the power spectral shape during the soft-to-hard state transition. In \citet{2015MNRAS.452.3666S}, we found that for observations of \gx339\ and other black hole X-ray binaries taken in the LHS during outburst rise, there is at least one BLN component which has a smaller characteristic frequency in the 1 -- 2 keV band compared to its characteristic frequency in the 2 -- 10 keV band. In this study, we find this difference in characteristic frequency only for obs.\ 4 and 5, while for all other observations the characteristic frequencies in these two bands agree within errors. According to the picture given in \citet{2015MNRAS.452.3666S}, characteristic frequencies, which are consistent between both bands, imply that the disk ends far away from the black hole and that the photons in both bands experience a similar number of scatterings. The decrease of the break frequency in the PDS along outburst decay, indicates an increasing inner disk radius \citep{1999A&A...352..182G}.     

Investigation of the combined \xmm/\nus\ energy spectra revealed a decreasing photon index, which is in agreement with the hardening of \gx339\ during outburst decay. The evolution of the accretion disk parameters (decreasing disk temperature, increasing inner disk radius) is in favour of the truncation disk model \citep[\eg][]{2001A&A...373..251D,2006csxs.book..157M,2007A&ARv..15....1D}. Regarding the inner disk radius obtained from the reflection component we find that the radius stays around $20\pm5\ R_g$ in the first four observation and then increases dramatically to over 75 $R_g$, assuming a neutral reflection disk. Given the large inner disk radius, which indicates a truncated accretion disk, a neutral disk is to be expected. If we allow for a free ionisation parameter, we find a highly ionised disk in the first observation, and a neutral disk in obs.\ 2, 3, and 4. Assuming a highly ionised disk in the first observation results in a smaller inner disk radius (and also in changes in the other spectral parameters) in this observation. The evolution of the spectral parameters is still consistent and in agreement with the truncation disk model.

The discrepancy between the inner disk radius of the accretion disk and that of the reflector can be related to the assumptions made in each model. The exact value of the inner disk radius derived from the disk blackbody normalization depends on the value of the hardening factor used (here 1.7). The uncertainty in the value of the hardening factor should not affect the evolution of the inner disk radius, as long as the hardening factor does not evolve during outburst decay. The \texttt{diskbb} model neglects the zero-torque inner boundary condition at the ISCO \citep{2005ApJ...618..832Z} and the effects of strong disk irradiation by the hard X-rays \citep{2008MNRAS.388..753G,2016A&A...590A.132V}. The later effect should only be important in the first (few) observations, as both model components indicate a disk truncated far away from the ISCO in the last two observations. The value of the inner radius obtained from the reflection component depends on the ionisation state and surface density structure of the disk and on disk inclination \citep{2014MNRAS.439.2307F}. Furthermore, there is a degeneracy between the spin and the inner accretion disk, as for current available data, decreasing the spin or increasing the inner edge of the disk can be considered to be almost equivalent. In addition, currently available reflection models describe the illumination of an otherwise cold slab of gas. They do not take into account the hotter surface layers of the disk in the case of a black hole X-ray binary, which will have a significant effect upon the reflection spectrum \citep{2007MNRAS.381.1697R}. Effects of the spatial extent of the corona, of coronal elevation and of mild relativistic outflow on the reflection spectrum, are also not included in currently available reflection models.
In contrast to spectral studies of \gx339\ during outburst rise based on \swift\ and \nus\ observations \citep{2015ApJ...808..122F,2016ApJ...821L...6P}, our spectra do not show indications of a broad iron line, and we do not require two different photon indices for the Comptonisation and reflection component.

The \xmm\ dataset allowed us to study for the first time the evolution of covariance spectra and ratios during outburst decay. Up-to-now studies of these properties mainly focus on outburst rise \citep{2009MNRAS.397..666W,2015MNRAS.452.3666S}. Our study reveals that while the source hardens, the energy-averaged covariance ratio increases. This increase shows that the variability on long time scales (low frequencies) contributes more to the overall variability compared to the variability at short time scales (high frequencies) while \gx339\ evolves along outburst decay. This behaviour also shows up in the PDS where the power in the top-flat part of the PDS at frequencies below 0.04 Hz clearly increases while the power in the decaying part of the PDS at frequencies above $\sim$0.3 Hz shows much less evolution, while the source hardens. 

Furthermore, we find that the increase in covariance ratio towards lower energies steepens, while \gx339\ hardens. The observed steepening of the increase implies that additional variability on long time scales becomes more and more important at soft energies while the source gets harder. In the scenario, in which the additional variability on long time scales and soft energies is thought to be due to intrinsic instabilities in the accretion disk \citep{2009MNRAS.397..666W}, which can be invoked by damped mass accretion rate variations or oscillations in the disk truncation radius \citep{1997MNRAS.292..679L,2003A&A...402.1013M}, the observed steepening of the increase indicates an increase in the disk instabilities when \gx339\ hardens. The observed evolution suggests that the stable disk of the soft state develops instabilities which get stronger when the source hardens.   
  
\acknowledgments
We thank the referee for thoughtful comments that helped to improve the clarity of our paper.
This project is supported by the Ministry of Science and Technology of
the Republic of China (Taiwan) through grants 104-281-M-007-060, 105-2112-M-007-033-MY2 and 105-2811-M-007-065.
Based on observations obtained with \xmm, an ESA science mission with instruments and contributions directly funded by ESA Member States and NASA.
This research has made use of data obtained through the High Energy Astrophysics Science Archive Research Center Online Service, provided by the NASA/Goddard Space Flight Center.

{\it Facilities:} \facility{\xmm}, \facility{\nus}, \facility{\swift}.

\bibliographystyle{apj}
\bibliography{/mybib}




\end{document}